\documentstyle[epsfig]{elsart}

\makeatletter
\if@twoside
\else \oddsidemargin 00pt \evensidemargin 00pt
\fi
\let\@eqnsel = \hfil

\expandafter\ifx\csname mathrm\endcsname\relax\def\mathrm#1{{\rm #1}}\fi
\@ifundefined{mathrm}{\def\mathrm#1{{\rm #1}}}{\relax}

\makeatother

\begin{document}

\begin{frontmatter}
\title {{\bf  Chargino Production at an $e^- e^-$ Collider  }}

\author[ad2]{J. Gluza}
and
\author[ad1]{T. W\"ohrmann}
\address[ad2]{ Department of Field Theory and Particle Physics, University of
Silesia, Uniwersytecka 4, PL-40-007 Katowice, Poland, e-mail:
gluza@us.edu.pl}
\address[ad1]{Institut f\"ur Theoretische Physik, Universit\"at W\"urzburg,
D-97074 W\"urzburg, Germany, e-mail: woerman@physik.uni-wuerzburg.de}
\begin{abstract}
The chargino pair production in $e^-e^-$ collisions with their subsequent
decays are considered within SUSY models with R-parity violation and with
lepton number non-conservation. The production process ($\sqrt{s}=1$ TeV) is
predicted to be large in a wide range of both sneutrino and chargino masses.
The influence of all virtual sneutrino states and their mixings with electrons
are taken into account. Some specific situations are pointed out when
significant suppressions of the cross section can take place.
The chargino decays are discussed for either the chargino
as LSP or the chargino as heavier sparticle.
In both cases unique signals are possible with
up to six charged fermions and without missing energy.
\end{abstract}
\end{frontmatter}

\section{Introduction}
Many problems connected with
particle physics as well as with astrophysics and cosmology remain unsolved
within the framework of the Standard Model (SM).
The search for
going beyond the SM is therefore a part of many research programs.
Supersymmetric models are among them as one of the most attractive
extensions of the SM.

In the Minimal Supersymmetric Standard Model (MSSM) we assume that lepton-
as well as baryon-numbers are conserved by introducing the so called R-parity
in order to avoid rapid proton decay. This number is given by
$R=(-1)^{L+3B+2S}$, with lepton number $L$, baryon number $B$ and
spin $S$ of the respective particle.
Two important
consequences arise by that: firstly, we may only produce an even number
of supersymmetric particles and, secondly, the lightest supersymmetric particle
(LSP) is stable (and it has to be the lightest neutralino $\tilde \chi_1^0$).
But there is no fundamental principle to introduce this R-parity.
As long as either the baryon- or the lepton-number remains conserved, we are
free to violate R-parity. In models with broken R-parity the
phenomenology will be changed significantly. Since the LSP is unstable
in these models, events with missing energy are not typical, anymore.
Therefore it is worthwhile to
investigate these models more carefully. It was done lately especially
in the context of the recent HERA data \cite{zeus} which can (also) be
interpreted within this type of models \cite{supdesy}.
In this paper we would like to consider possible R-parity violation
signals from
chargino production at an $e^-e^-$ collider. This option of a future linear
collider is especially promising for non-standard physics as the SM
activity is highly suppressed (doubly charged initial state with a finite
lepton number) \cite{nlc}.

Since this process violates lepton number, we introduce an additional
superpotential violating R-parity, allowed by supersymmetry and
renormalization \cite{aulakh}.
This superpotential leads to new couplings both between
two leptons and a slepton and between two baryonic states and
a leptonic one
\begin{eqnarray} \cal{L} &=&
\lambda_{abc} \{\tilde \nu_{aL} \bar e_{cR} e_{bL} + \tilde e_{bL}
\bar e_{cR} \nu_{aL} + (\tilde e_{cR})^{\ast }
(\bar \nu_{aL})^c e_{bL} -(a \leftrightarrow b )\} + h.c.\\
\cal{L} &=&
\lambda^{\prime }_{abc} \{\tilde \nu_{aL} \bar d_{cR} d_{bL} + \tilde d_{bL}
\bar d_{cR} \nu_{aL} + (\tilde d_{cR})^{\ast } (\bar \nu_{aL})^c d_{bL} -
\nonumber  \\ & &  \tilde e_{aL} \bar d_{cR} u_{bL} - \tilde u_{bL}
\bar d_{cR} e_{aL} + (\tilde d_{cR})^{\ast } (\bar e_{aL})^c u_{bL} \} + h.c.
\end{eqnarray}
We also get additional mixing between leptons, gauginos and higgsinos
$( e, \mu , \tau , \tilde W^- , \tilde H^- )$ and
$( \nu_e , \nu_{\mu } , \nu_{\tau } , \tilde \gamma , \tilde Z ,
\tilde H_1^0 , \tilde H_2^0 )$, charged sleptons and  charged Higgs
bosons $( \tilde e , \tilde \mu , \tilde \tau , H^- )$ and sneutrinos
and neutral Higgs bosons $( \tilde \nu_e , \tilde \nu_{\mu },
\tilde \nu_{\tau }, H_1^0 , H_2^0 )$
(for details of these mixings see e.g. \cite{bartl}).
From experiments we know, that
these additional couplings, though existing, are small compared to the gauge
ones \cite{bhat}. We have also strong restrictions for additional mixings
except those between gauginos and higgsinos, and between two neutral
Higgs states.

The number of processes, in which two particles are produced is, due to
charge conservation, limited to
\begin{enumerate}
\item $ e^- e^- \rightarrow e^- e^- $
\item $ e^- e^- \rightarrow W^- W^- $
\item $ e^- e^- \rightarrow l^-_i l^-_j \qquad  l_{i(j)}=\mu^-,\tau^- $
\item $ e^- e^- \rightarrow \tilde e^- \tilde e^- $
\item $ e^- e^- \rightarrow \tilde l^-_i \tilde l^-_j \qquad
\tilde{l}^-_{i(j)}=\tilde{\mu}^-,\tilde{\tau}^- $
\item $ e^- e^- \rightarrow \tilde \chi^-_i \tilde \chi^-_j \qquad i,j=1,2 $
\end{enumerate}
Process (i) is the only possible process allowed in the SM and already
investigated \cite{eeee}. Processes (ii) and (iii) violate lepton number,
but they are
also possible in models without supersymmetry and have been investigated
\cite{eeww}.
Processes (iv)-(vi) are
only possible in supersymmetric models, where process (iv) is the only one,
which is also possible within the MSSM, because it does not violate
lepton number. This process has also been
considered in the literature \cite{cuyp}.
Both processes (v) and (vi) violate lepton number. Since
process (v) is limited by the couplings $\lambda_{abc}$, we will
consider only process (vi) in this paper. Although this process was already
discussed in \cite{cuyp}, both sneutrino
mixing effects and R-parity violating decays of the produced charginos
were not taken into account.
It was assumed that the LSP is still the lightest neutralino and stable
(like in the MSSM with conserved both R-parity and lepton number).
Consequently the signatures of the final states are quite different from those
given there.

This paper is organized as follows: in the next section we will calculate
and discuss the cross section for process (vi) and in the following section
we shall investigate possible decay modes of the produced charginos
and we will consider possible signatures. Some concluding remarks will be
collected in the last section.

\section{Chargino production by $e^-e^-$ scattering}

The process $e^-e^- \rightarrow \tilde{\chi}_i^- \tilde{\chi}_j^-$ proceeds
through two Feynman diagrams with exchanged sneutrinos (in t and u channels)
so the sneutrino-chargino-electron coupling must be known.
In supersymmetric models the charginos are in principle mixtures
of a Wino and a charged Higgsino eigenstate. If the lepton number
is violated, then also leptons may mix with the Wino and the Higgsino.
However, as already mentioned before, this mixing is strongly restricted by
experiments and therefore
negligible for the considered production process. The remaining mixing may be
written in the following way $ \tilde \chi^-_i = V_{i 1} \tilde W^- + V_{i 2 } \tilde H^- $.
Since only the Wino contribution couples with an electron (in the limit of
zero electron mass) we will assume normalization in which the produced chargino
is a pure Wino state with
$V_{i1}=1$.
For other mixings results given in this
paper should be multiplied by a factor $V_{i1}^{4}$.

This assumption implies the chargino-sneutrino-electron gauge coupling to be
\begin{equation}
{\cal L}=\frac{g}{\sqrt{2}}
\bar{\tilde{\chi}}^c_i(1-\gamma_5)e\tilde{K}_{em}\tilde{\nu}_{m}^{\ast}
+ h.c.,
\end{equation}
with $g$ defined by the SM charged Lagrangian
\begin{equation}
{\cal L}=\frac{g}{\sqrt{2}}
\bar{e}\gamma^{\mu}(1-\gamma_5)K_{em}{\nu}_m W_{\mu}^-+ h.c..
\end{equation}

These interactions are written in a physical basis.
For simplicity we assume that only left-handed currents exist.
If we do not introduce right-handed neutrino $SU(2)_L$ singlets to the SM
then neutrinos are massless and $K=I$, otherwise we have a $(3+n_R) \times
(3+n_R)$ neutrino mass matrix and consequently a $3 \times (3+n_R)$ charged
lepton-neutrino mixing matrix $K$ which is of course not equal to identity
$K \neq I$ (for details see e.g. \cite{glzr}).
Similarly a diagonalization of the  sneutrino - neutral Higgs sector gives
a $\tilde{K}$ mixing matrix of dimension $3 \times 5$\footnote{To be more
specific: Let's denote sneutrino and neutral Higgs states by
$\tilde{\nu}_i$,
i=1,...,5. Then introducing a unitary $5 \times 5$ matrix $\tilde{U}=\left(
\matrix{
\tilde{K} \cr \tilde{K}' } \right)$, which diagonalizes the sneutrino-neutral
Higgs  mass matrix, we get also a transformation between physical and
non-physical states $\tilde{\nu}=\tilde{U}\tilde{\nu}^{phys.}$. Assuming
that the neutral Higgs coupling with the chargino-electron pair is negligible,
we get directly the matrix $\tilde{K}$ in Eq.(3).}. 
Due to unitarity of the matrix $\tilde{U}$ we have
\begin{equation}
\sum_{m=1}^{5} \mid \tilde{K}_{em} \mid^2=1.
\end{equation}

In the following, when we become specific,
we will only take into account the sneutrino
mixing submatrix with dimension $3\times 3$. This is mainly motivated
by the fact that in the limit of zero electron mass only sneutrino-electron
couplings are unequal to zero. Both, the Higgs mixing part and the sneutrino
Higgs mixing part are of no interest, since the first does not lead to any
coupling (this is also the reason why an appropriate part of the Lagrangian
with the $\tilde{K}'$ submatrix does not appear in Eq.(3)) and the second
is assumed to be small.

In the limit of zero electron mass only four nonzero  helicity amplitudes
 exist which can be written in the following
form ($\lambda_{i(j)}=\pm 1/2$ stands for the helicities of the produced
charginos, $\Sigma\lambda=
\lambda_i+\lambda_j$, $\Theta$ is a scattering  angle of the chargino
$\tilde \chi^-_i$ in the $e^-e^-$ CM energy frame)
\begin{eqnarray}
M(--;\lambda_i,\lambda_j)&=&\frac{g^2}{2}\sum_{m=1}^{3} \tilde{K}_{em}^2 s
\sqrt{(E_i+2\lambda_ip_i)
(E_j+2\lambda_jp_j)} \nonumber \\
&& \left(\frac{1}{t-m_{\tilde{\nu}_m}^2}-
\frac{1}{u-m_{\tilde{\nu}_m}^2}\right)D^1_{-1,\Sigma\lambda}(\Theta),
\end{eqnarray}
where
\begin{eqnarray*}
E_{i(j)}&=&\frac{s \pm m_{\tilde{\chi}_i^-}^2\mp m_{\tilde{\chi}_j^-}^2}
{2\sqrt{s}}, \\
t(u)&=&m_{\tilde \chi^-_i}^2-\sqrt{s}(E_i \mp 2 p_i \cos{\Theta}).
\end{eqnarray*}

Two factors influence the magnitude of the amplitude $M$ -- the square of
the mixing matrix elements $\tilde{K}_{me}^2$ and the sneutrino masses
$m_{\tilde{\nu}_m}$. The interplay between different elements of
the $\tilde{K}_{em}$ matrix and the sneutrino masses can be quite important
and can impact on the magnitude of the cross section in a meaning way as we
will discuss now.

Let's assume that all sneutrinos are degenerate in mass and
the matrix $\tilde{K}$ is real.
Then the amplitude $M$ in Eq.(6) has the form
\[ M \propto \sum\limits_{m} \tilde{K}_{em}^2
f(m_{\tilde{\nu}_m}) = f(m_{\tilde{\nu}_m}) \sum\limits_{m}
\tilde{K}_{em}^2 = f(m_{\tilde{\nu}_m}),\]
where Eq.(5) has been used.
That means that in this case, no matter whatever the
mixing between the sneutrinos and the electrons is, we have effectively
one sneutrino
with maximal mixing $\tilde{K}_{em}=1$. A similar situation happens when
we have different
masses of the sneutrinos but only with one dominate (maximal) mixing
$\tilde{K} _{em} \simeq 1$ (negligible contributions from the other ones).

Fig.1 shows results for these two cases when a pair of the same chargino is
produced at CM energy equal to 1~TeV.
The results are also sensitive to the free parameters of the
neutralino-chargino sector. Instead of choosing parameters
$M_{\frac{1}{2}}$, $\mu$ and $\tan \beta = \frac{v_2}{v_1}$ as it is
usually done, we fix here the chargino parameters
$200\;\mbox{GeV} \leq m_{\tilde \chi_1^- } \leq 500\;\mbox{GeV} $
and $V_{11}=1$. Notice that this choice is always possible, since the
chargino sector is described by three free parameters, but for our calculations
we have to fix only two of them. The sneutrino mass is taken in the range
$100\;\mbox{GeV} \leq m_{\tilde{\nu}} \leq 800$ GeV.

\vspace{9.5 cm}
\begin{figure}[h]
\includegraphics{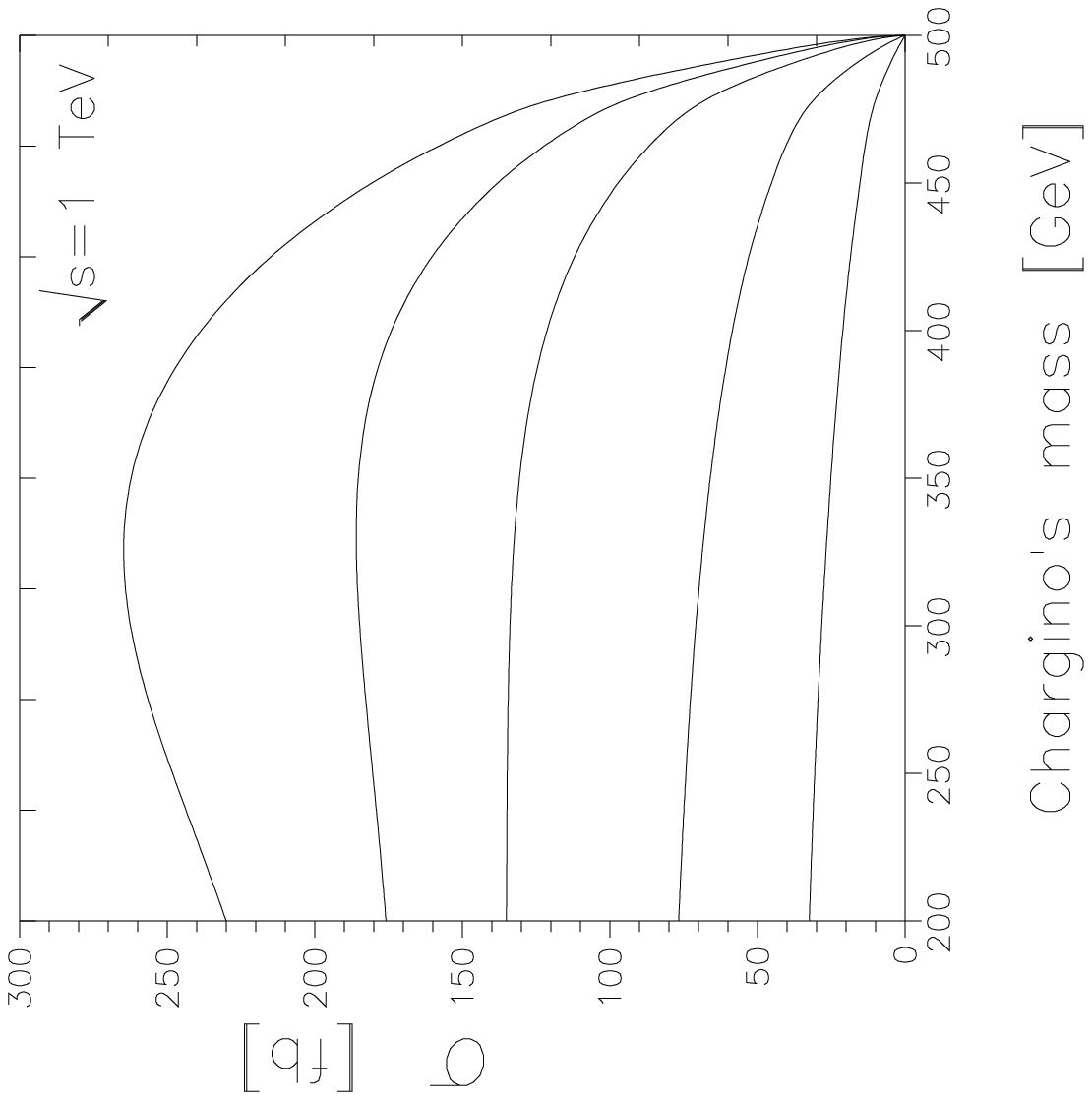}
\vspace{0.3 cm}
\end{figure}

\baselineskip 5 mm

{\footnotesize Fig.1
Cross section for the process
$e^-e^- \rightarrow \tilde \chi^-_1 \tilde \chi^-_1$
as a function of the chargino mass and sneutrino masses
(downwards) $m_{\tilde \nu}=100,200,300,500,800$ GeV and
$\sqrt{s}=1$ TeV, $\tilde{K}_{em}=1$, $V_{11}=1$ (see text for details).}

\vspace{0.5 cm}

We can see that for a large spectrum of both sneutrino and chargino
masses significant (detectable) values of the cross section are possible.

But a different situation can arise when the
mixing matrix element $\tilde{K}_{em}$ is not the maximal one and consequently
smaller cross sections will be achieved
($\sigma(e^-e^- \rightarrow \tilde \chi^-_1 \tilde \chi^-_1) \sim
\tilde{K}_{em}^4$).
Firstly, let's assume that the
sneutrinos have apparently different masses but the mixing elements have
comparable values.
Such a situation is possible. To make this point clearer let's
support this statement by recalling an example from neutrino physics.
It was proposed lately that
solar, atmospheric and LSND results can be explained with a
neutrino mixing matrix of the form \cite{ackpak} (notations modified
to our convention (Eq.(4)), l(m) stands for charged leptons (neutrinos),
respectively)
\begin{equation}
\left( \matrix{ .700 & .700& .140 \cr
                          -.714 & .689 & .124 \cr
                           -.010 & -.187 & .982 } \right)
\leq K_{lm} \leq
\left( \matrix{ .764 & .630& .140 \cr
                          -.645 & .754 & .124 \cr
                           -.028 & -.185 & .982 } \right)
\end{equation}

Similarly, if some of the elements of the matrix $\tilde{K}$ are
comparable in magnitude then,
because of kinematical reasons, only the lightest sneutrino would
contribute to the cross section (see Eq.(6)) and there would be a
possibility of results smaller by even one order of
magnitude (e.g. $\sim {(0.7)}^4$)
compared with the situation in Fig.1.
Secondly, until now we assume that the $\tilde{K}_{em}$ elements are real.
However, if some of them are imaginary
then real and imaginary elements
would give opposite contributions to
the amplitude (Eq.(6)) causing that the cross section would also be smaller
than that given in Fig.1. We could even imagine some dramatic cancelations
in this case. Let's invoke physics from the neutrino sector once more.
Another possible solution to a few not fully understood neutrino
experimental results have been postulated e.g. in \cite{mohnus}.
It was found
that almost degenerate three neutrinos can reconcile solar, atmospheric,
dark matter  and
neutrinoless double-$\beta$ decay results with a mixing matrix of the form
\begin{equation}
K_{lm}= \frac{1}{\sqrt{3}} \left( \matrix{ 1 & \omega & \omega^2 \cr
                                           1 & \omega^2 & \omega \cr
                                           1 & 1 & 1 } \right),
\quad (\omega=e^{\frac{2\pi i}{3}}).
\end{equation}

For such a matrix we have
$\sum\limits_m {K_{em}^2}=1+\omega+\omega^2=0$. If a similar situation exists
in the sneutrino sector then a large suppression of the cross section could
happen.

\section{Decays and Signatures}
In this section we shall discuss possible decays of the produced charginos
in order to investigate possible signatures for the process
$e^- e^- \rightarrow \tilde \chi_1^- \tilde \chi^-_1$. Since
the lightest supersymmetric particle LSP is in case of R-parity
violation not necessarily
the lightest neutralino $\tilde \chi_1^0$ but also the lighter chargino or
the sneutrino $\tilde \nu$
are possible candidates, we have to distinguish two possible scenarios:
\begin{enumerate}
\item $\tilde \chi^-_1 $ is the LSP
\item $\tilde \chi^-_1 $ is not the LSP
\end{enumerate}
\begin{enumerate}
\item In case the chargino is the LSP, it decays R-parity violating.
Two different modes are possible: In the first case all lepton components of
$\tilde \chi_1^- $ are negligible.
Then the chargino decays
via t- and u-channel processes with a virtual sfermion
\[
\tilde \chi_1^- \rightarrow e_i^- \nu_j \bar \nu_k \, (e_i^+ e_j^- e_k^-)
\]
or
\[
\tilde \chi_1^- \rightarrow e_i^- q_j \bar q_k \, (\nu_i \bar u_j d_k)
\]
due to the couplings $\lambda_{abc}$ or  $\lambda'_{abc}$ at the
sfermion-fermion-fermion vertex
respectively, where the $i,j,k=1,2,3$ denote the families.

In the second case a non-vanishing lepton-wino and lepton-higgsino
mixing exists.
This mixing, although (as already discussed) negligible for the production
process, may become
important for the decays \cite{bartl}. In this case, the lepton component of
$\tilde \chi_1^-$ couples to a gauge boson and the respective lepton by a
gauge coupling. Therefore the interesting decay modes are
\[
\tilde \chi_1^- \rightarrow W^- \nu_i \, (Z e^-_i)\, (i=1,2,3).
\]
Therefore possible signatures contain up to six charged fermions, in
some cases six charged leptons. In each case signatures without
missing energy are possible.
\item In case the lighter chargino $\tilde \chi_1^-$ is not the LSP, we
have to consider other candidates. Two possible candidates, discussed here,
are the lightest neutralino $\tilde \chi_1^0$ or the sneutrino $\tilde
\nu$.

In some scenarios the chargino, though not the LSP, still decays
dominantly R-parity violating, into either three fermions \cite{dreiner} or
a fermion and a gauge boson \cite{bartl}. For these scenarios the modes
and therefore the signatures are just the same as in case (i).

In many other scenarios the chargino decays R-parity conserving
into the
respective LSP and SM-particles \cite{cuyp}

\begin{eqnarray}
LSP=neutralino && \nonumber \\
\tilde \chi_1^- & \rightarrow & \tilde \chi_1^0 \bar \nu_l e_l^-
(\tilde{\chi}_1^0 \bar u_i d_i) \nonumber \\
\tilde{\chi}_1^-  & \rightarrow & H^-\tilde{\chi}_1^0, \nonumber \\
\tilde{\chi}_1^-  & \rightarrow & \, W^- \tilde{\chi}_1^0 \nonumber \\
LSP=sneutrino \nonumber \\
\tilde \chi_1^- & \rightarrow & \tilde \nu_l e_l^-. \nonumber
\end{eqnarray}
The respective LSP subsequently decays R-parity violating.
In the first case the $\tilde \chi_1^0 $ decays by $\tilde \chi_1^0
\rightarrow e_i^+ e_j^- \nu_k
(e_i^- u_j \bar{d}_k)$ or, if there is a non-vanishing
neutrino-gaugino-higgsino
mixing, also the decays $\tilde \chi_1^0 \rightarrow \nu_i Z\,
(e_i W)$ are possible. In the latter case the sneutrino $\tilde \nu_i$
decays by $\tilde \nu_i \rightarrow e^+_j e^-_k (\bar q_j q_k)$.
\end{enumerate}

The main result is that, for all different scenarios, signatures without
missing energy are possible. Since missing energy is a typical part of a
MSSM event, these signatures are obviously different from those for the
process $e^- e^- \rightarrow \tilde e^- \tilde e^-$ within the MSSM.

Backgrounds from SM or non-SUSY models with lepton number violation are of
higher order, since the basic processes $e^- e^- \rightarrow e^- e^- $
and $ e^- e^- \rightarrow W^- W^- $ lead to signatures with
at most four charged fermions. Therefore possible background processes
are $e^- e^- \rightarrow l l BB$, where $l$ denotes a lepton and
$B$ denotes a boson
$B=H,W,Z$ \cite{back}. These higher order processes should be compared
with signatures considered in this paper as both of them predict 6
charged fermions in the final state (paper in preparation).

\section{Conclusions}

We have discussed chargino pair production in $e^-e^-$ collisions for
supersymmetric models with R-parity violation and with lepton number
non-conservation. The gauge nature of the appropriate couplings causes that
the magnitude of the cross section is large enough for the process to be
detected in a wide range of both sneutrino and chargino masses. However,
it has been shown that the interplay between contributions from
different sneutrino
states can cause large suppression of the cross section. If this is not the
case then the final signal from the decaying charginos is worth to study. It
has been shown that final state with up to six charged fermions or even, in
some cases, six charged leptons can be produced for both
considered cases, e.g. the lighter chargino is the LSP or is not the LSP.
These signatures are without missing energy.

We therefore conclude, that an $e^- e^-$-collider can be a very good tool
in order to discover SUSY models with lepton number violation.

\section*{Acknowledgments}

T.W. is partly supported by the German Federal Ministry for Research and
Technology (BMBF) under contract number 05~7WZ91P~(0), J.G.
by the Polish Committee for Scientific Research, Grant No. PB659/P03/95/08
and the University of Silesia internal Grant. J.G. appreciates also the
fellowship of the Foundation for Polish Science.

We are grateful to H.~Fraas, S.~Hesselbach
and M.~Zra{\l}ek for many helpful discussions.

Finally, T.W.
likes to thank the Department of Field Theory and Particle Physics of
the University of Silesia in Katowice for the kind hospitality
during his stay, when part of this project was done.

\end{document}